\documentclass[12pt]{article}

\usepackage{amsmath,amssymb,amsthm,epsf,epsfig}
\usepackage{amsmath,amssymb}

\begin{document}
\begin{center}
\large{\textbf{\textbf{RELATIVISTIC THERMODYNAMICS WITH AN INVARIANT ENERGY SCALE}}}\\
\end{center}
\begin{center}
Sudipta Das$^{a,}$\footnote{E-mail: sudipta\_jumaths@yahoo.co.in},
Subir Ghosh$^{a,}$\footnote{E-mail: sghosh@isical.ac.in} and
Dibakar Roychowdhury$^{b,}$\footnote{E-mail: dibakar\_nbu@yahoo.co.in} \\
$^a$Physics and Applied Mathematics Unit, Indian Statistical
Institute\\ 203 B.T.Road, Kolkata 700108, India \\
$^b$Department of Physics, University of North Bengal \\ Siliguri
734013, West Bengal, India
\end{center}
{\textbf{Abstract:}}~~~ A particular framework for Quantum Gravity
is the Doubly Special Relativity (DSR) formalism that introduces a
new observer independent scale, the Planck energy. Our aim in this
paper is to study the effects of this energy upper bound in
relativistic thermodynamics. We have explicitly computed the
modified equation of state for an ideal fluid in the DSR
framework. In deriving our result we exploited the scheme of
treating DSR as a  non-linear representation of the Lorentz group
in Special Relativity.

\section {\bf {Introduction}}
In recent years Doubly Special Relativity (DSR) \cite{ga} has
created a lot of interest as a possible framework of Quantum
Gravity. This is mainly due to two basic tenets on which the
theory rests: (i) Appearance of a second observer independent
scale \cite{ga}, which can be (Plank) length, (Planck) energy or
momentum, apart from the velocity of light, common to Special
Relativity (SR). Incidentally this gives rise to the name DSR.
(ii) A naturally emerging Non-Commutative (NC) spacetime \cite{ga,
kowal, sg, rev} {\footnote{NC geometry in general has generated a
lot of new ideas in modern physics \cite{sw}}}. Both of these
features are very close to Quantum Gravity ideas \cite{planck} or
the existence of a universal short distance scale that postulates
a generalized uncertainty principle \cite{gup}. All the models of
Quantum Gravity predict qualitatively different spacetime beyond
certain energy (length) scale, generally considered to be the
Planck energy (length). Also it is now established \cite{dfr} that
a consistent marriage of ideas of quantum mechanics and
gravitation requires NC spacetime to avoid the paradoxical
situation of creation of a black hole for an event that is
sufficiently localized in spacetime. Quite obviously one would
like to have the numerical value of this universal scale to be
observer independent, as in DSR.

In this perspective it is very important to study the effects of
specific forms of NC spacetime that are relevant to DSR, in
particular the $\kappa$-Minkowski spacetime, studied independently
\cite{maj} and partly motivated by DSR ideas \cite{ga, kowal, luk,
ms}. So far only particle dynamics in DSR framework has been
studied, which has revealed many unusual features \cite{kowal,
mig}. Some field theory models in DSR spacetime have also been
attempted \cite{fil}. On the other hand, to our knowledge not much
work has been done in studying DSR effects (especially the fact
that there exists an upper bound of energy) in the exciting areas
of relativistic thermodynamics and eventually in cosmology. In the
present paper we have initiated a study along the direction of
relativistic thermodynamics in the DSR framework.

Our aim is to follow the prescription of Weinberg \cite{wein}
where one postulates an explicit form of the Energy-Momentum
Tensor (EMT) for a perfect fluid in the Lagrangian framework. The
first non-trivial task that we face is the construction of the
DSR-covariant EMT. Fortunately we have a powerful tool at our
disposal: {\it{DSR kinematics is a manifestation of a non-linear
realization of SR kinematics}} \cite{ms, visser, sg}. Throughout
the present paper we exploit this principle to develop the fluid
EMT for DSR and subsequently study the consequences of the EMT in
thermodynamic context. As expected our expressions have a smooth
commutative (or equivalently SR) limit, that is, all results reduce
to SR results when $\kappa$, the effective NC parameter (the
energy upper bound) goes to infinity.

The paper is organized as follows: In Section 2 we will provide
the explicit non-linear mapping between the NC DSR variables
(expressed as small letters) and commuting (or more precisely
canonical) degrees of freedom (expressed as capital letters). The
latter obey canonical phase space algebra and SR Lorentz transformations
whereas the former satisfy NC $\kappa$-Minkowski phase space
algebra and DSR Lorentz transformations. In Section 3 we will
construct the DSR compatible EMT. This is one of our main results.
In Section 4 we will explicitly reveal effects of DSR
regarding relativistic thermodynamics which constitutes the
other major result. We will conclude in Section 5.

\section {\bf {Nonlinear Realization of Lorentz Group}}
It has been pointed out by Amelino-Camelia \cite{ga} that there is
a connection between the appearance of an observer independent
scale and the presence of non-linearity in the corresponding
spacetime transformations. Recall that Galilean transformations
are completely linear and there are no observer independent
parameters in Galilean/Newtonian relativity. With Einstein
relativity one finds an observer independent scale - the velocity
of light - as well as a non-linear relation in the velocity
addition theorem. In DSR one introduces another observer
independent parameter, Planck energy or length, and ushers another
level of non-linearity in which the Lorentz transformation laws
become non-linear. These generalized Lorentz transformation rules,
referred to here as DSR Lorentz transformation, are derivable from
basic DSR ideas \cite{ga} or in a more systematic way, from
integrating small DSR transformations in an NC spacetime scheme
\cite{br, sg}. Another elegant way of derivation is to interpret
DSR laws as a non-linear realization of SR laws where one can
directly exploit the non-linear map and its inverse, that connects
DSR to SR and vice-versa {\footnote{It needs to be stressed that
even though there exists an explicit map between SR and DSR
variables, the two theories will not lead to the same physics, (in
particular upon quantization), due to the essential non-linearity
involved in the map. According to DSR the physical degrees of
freedom live in a non-canonical phase space and the canonically
mapped phase space is to be used only as a convenient intermediate
step.}}. Obviously to accomplish this one needs the map, which can
be constructed by a motivated guess \cite{ms,visser} or
constructed as a form of Darboux map \cite{sg}.

We are working in the DSR2 model of Magueijo and Smolin \cite{ms}.
Let us start with the all important map \cite{ms, visser, sg},
\begin{equation}
 F(X^{\mu})\rightarrow x^{\mu}~~,~~ F^{-1}(x^{\mu})\rightarrow X^{\mu} \label{1}.
\end{equation}
which in explicit form reads:
\begin{equation}
F(X^\mu)=x^{\mu}\left(1-\frac{(n p)}{\kappa}\right)~~~;~~~
F(P^\mu)=\frac{p^{\mu}}{\left(1-\frac{(n p)}{\kappa}\right)}
$$$$F^{-1}(x^{\mu})=X^{\mu}\left(1+\frac{(n P)}{\kappa}\right)~~~;~~~
F^{-1}(p^{\mu})=\frac{P^{\mu}}{\left(1+\frac{(n
P)}{\kappa}\right)} \label{ld}
\end{equation}
where $n_{\mu}=(1, 0, 0, 0)$ and we use the notation $a_\mu
b^\mu=(ab),~(n p)=p^0,~(n P)=P^0$. Note that upper case and lower
case letters refer to (unphysical) canonical SR variables and
(physical) DSR variables respectively.

As a quick recapitulation let us rederive the DSR Lorentz
transformations ($L_{DSR}$), starting from the familiar (linear)
SR Lorentz transformations ($L_{SR}$),
\begin{equation}
X'^0=L_{SR}(X^0)=\gamma(X^0-vX^1)~~,~~X'^1=L_{SR}(X^1)=\gamma(X^1-vX^0)~~,~~X'^2=X^2~~,
~~X'^3=X^3
$$$$ P'^0=L_{SR}(P^0)=\gamma(P^0-VP^1)~~,~~P'^1=L_{SR}(P^1)=\gamma(P^1-VP^0)~~,
~~P'^2=P^2~~,~~P'^3=P^3 \label{lcxp}
\end{equation}
where $\gamma =1/\sqrt{1-v^2}$ and the boost is along $X^1$
direction with velocity $v^i = (v, 0, 0)$. Note that  the second
line of (\ref{lcxp}) involves $V$ but following our definition
$\frac{dX^i}{dX^0}=\frac{dx^i}{dx^0}$ so that $V=v$. Now the DSR
Lorentz transformation $L_{DSR}$ is formally expressed as,
\begin{equation}
x'^\mu = L_{DSR} (x^\mu) = F \circ L_{SR} \circ F^{-1}
(x^\mu)~~~,~~~p'^\mu = L_{DSR} (p^\mu) = F \circ L_{SR} \circ
F^{-1} (p^\mu). \label{lnc}
\end{equation}
In explicit form this reads as,
\begin{equation}
x'^0 = L_{DSR} (x^0) = F \circ L_{SR} \circ F^{-1} (x^0) = F \circ
L_{SR} \left(X^0\left(1+\frac{P^0}{\kappa}\right)\right) $$$$ = F
\left(\gamma \left(X^0 - v X^1\right)\left(1 +
\frac{\gamma}{\kappa}\left(P^0 - vP^1\right)\right) \right) =
\gamma \alpha (x^0 - v x^1)~~; $$$$ p'^0 =
\frac{\gamma}{\alpha}(p^0 - v p^1) \label{lx0}
\end{equation}
where $\alpha=1+\kappa^{-1}\left(\left(\gamma-1\right)P^0-\gamma
vP^{1}\right)$. Similarly for $\mu=1$ we have the following
expressions:
\begin{equation}
x'^1 = \gamma \alpha (x^1 - v x^0)~~~,~~~p'^1 =
\frac{\gamma}{\alpha}(p^1 - v p^0). \label{lx1}
\end{equation}
{\it{It is important to realize that, in the present formulation,  noncommutative effects enter through these generalized
(non-linear) transformation rules.}}

Note that, in contrast to SR laws
(\ref{lcxp}), components of $x^\mu,~p^\mu$ {\it{transverse}} to
the frame velocity $v$ are affected,
\begin{equation}
x'^i = \alpha x^i~~~,~~~p'^i= \frac{p^i}{\alpha};~~i=2,3.
\end{equation}
There are two phase space quantities, invariant under DSR Lorentz
transformation:\\ $ \eta_{\mu \nu} p^\mu p^\nu / (1-p^0/\kappa)^2$
and $\eta_{\mu \nu} x^\mu x^\nu (1-p^0/\kappa)^2$ with
$\eta_{\mu\nu}=diag(-1, 1, 1, 1)$. Writing the former as
\begin{equation}
m^2=\eta_{\mu \nu} p^\mu p^\nu / (1-p^0/\kappa)^2 \label{ms}
\end{equation}
yields the well-known Magueijo-Smolin dispersion relation. We
interpret the latter invariant to provide an effective metric
$\tilde{\eta}_{\mu\nu}$ for DSR:
\begin{equation}
d \tau^2 = \tilde{\eta}_{\mu \nu} x^\mu x^\nu = (1-p^0/\kappa)^2
\eta_{\mu \nu} x^\mu x^\nu .\label{ncm}
\end{equation}

\section {\bf {Energy-momentum tensor in $\kappa$-Minkowski spacetime}}
In this section our aim is to construct the energy-momentum tensor
of a perfect fluid, that will be covariant in the DSR framework.
Indeed, this will fit nicely in our future programme of pursuing a
DSR based cosmology.
\subsection {\bf {Fluid in SR theory:}}~~~A perfect fluid can be considered
as a system of structureless point particles, experiencing only
spatially localized interactions among themselves. The idea is to
consider boosts in  a passive transformation framework. In this
way one can ascertain the structure of energy momentum tensor in
an arbitrary inertial frame (laboratory frame) by boosting the
expression valid in the fluid rest frame (comoving frame). In a
comoving Lorentz frame, from spherical symmetry, the energy
momentum tensor $\tilde T^{\mu\nu}$ of a perfect fluid becomes
diagonal and the components are explicitly written as,
\begin{equation}
\tilde{T}^{ii}=P ~~;~~\tilde{T}^{00}=D~~;~~\tilde T^{0i}=\tilde
T^{i0}=0. \label{lD0}
\end{equation}
The thermodynamic quantities $P$ and $D$ represent pressure and
energy density of the fluid. The components of the canonical
energy-momentum tensor transform under SR Lorentz transformation
$L_{SR}$ as a second rank tensor and in an arbitrary inertial
frame it assumes the form \cite{wein}
\begin{equation}
T^{00}=L_{c}(\tilde{T}^{00})=\gamma^2(D+Pv^2)~~,
~~T^{i0}=L_{c}(\tilde{T}^{i0})=\gamma^2(D+P)v^i $$$$
T^{ij}=L_{c}(\tilde{T}^{ij})=\gamma^2(D+P)v^iv^j + P \delta^{ij}.
\label{lct}
\end{equation}
The above set of equations can be integrated into a single SR
covariant tensor,
\begin{equation}
T^{\mu\nu}= (P+D)U^{\mu}U^{\nu}+ P\eta^{\mu\nu} \label{lctf}
\end{equation}
where the velocity $4$-vector $U^\mu$ is defined as
$U^0=\gamma~,~U^i=\gamma v^i$ with $U^\mu U_\mu=-1$.

We can derive this result in another way, the so called Lagrangian
formalism, which will be useful later. Let us treat the fluid as a
collection of non-interacting particles, the latter having in
general, an energy momentum tensor of the form \cite{wein}
\begin{equation}
T^{\mu\nu}=\sum_{i}\frac{P^{\mu}_{i}P^{\nu}_{i}}{P^0_{i}}
\delta^{3}\left(\bf{X} - \bf{X}_{i}\right) \label{lt1}
\end{equation}
where $P^{\mu}_{i}$ is the energy-momentum four-vector associated
with the $i$-th particle located at $\bf{X}_i$. Once again in the
comoving frame it will reduce to the diagonal form:
\begin{equation}
\tilde{T}^{ii}=P = \frac{1}{3}
\sum_{i}\frac{\bf{P}_{i}^{2}}{P^0_{i}}\delta^{3}\left( X - X_i
\right)~~;~~\tilde{T}^{00}=D =\sum_{i} P^0_i\delta^{3}\left(X -
X_i \right)~~; ~~\tilde{T}^{i0}=\tilde{T}^{0i}=0. \label{lD}
\end{equation}
In the above relations $P^0_i$ stands for the energy of the $i$-th
fluid particle. The thermodynamic quantities $P$ and $D$ represent
pressure and energy density of the fluid. The particle number
density is naturally defined as
\begin{equation}
N = \sum_{i}\delta^{3}\left( X - X_i\right) \label{lN}.
\end{equation}
The Lorentz transformation equation for $T^{\mu \nu}$ is
\begin{equation}
T^{\mu \nu}=L_{SR}(\tilde{T}^{\mu \nu})=\Lambda^\mu_\alpha
\Lambda^\nu_\beta T^{\alpha \beta}\label{n1}
\end{equation}
where $\Lambda$ is the Lorentz transformation matrix. For $\mu =
\nu = 0$ we have
\begin{equation}
T^{00}=(\Lambda^0_0)^2 \tilde{T}^{00} + (\Lambda^0_i)^2
\tilde{T}^{ii} = \gamma^2 \tilde{T}^{00} + \gamma^2 v^2
\tilde{T}^{11}. \label{n2}
\end{equation}
We put the summation expressions for $\tilde{T}^{00}$ and
$\tilde{T}^{ii}$ from (\ref{lD}) in the above equation instead of
putting $D$ and $P$. Then using the relation
\begin{equation}
T^\alpha_\alpha = -\sum_i\frac{m^2}{P^0_i}\delta^3(X-X_i) = -D+3P
\end{equation}
we can easily verify that the final expression for $T^{00}$ is exactly the same
as in (\ref{lct}). Similarly the other relations will follow.

\subsection {\bf {Fluid in DSR theory:}}~~~In order to derive
the expression for the DSR EMT ($t^{\mu\nu}$) we shall exploit the
same approach as above for SR EMT. Spatial rotational invariance
remains intact in DSR allowing us to postulate a similar diagonal
form for DSR EMT in the comoving frame. The next step (in
principle) is to apply the $L_{DSR}$ to obtain the general form of
EMT in DSR. We first define the non-linear mapping for the
energy-momentum tensor of a perfect fluid in a comoving frame. In
the second step we shall apply the Lorentz boost ($L_{SR}$) on our
mapped variable and finally arrive at the desired expression in
the DSR spacetime through an inverse mapping. But we will see that
when we try to introduce the fluid variables in the DSR EMT in
arbitrary frame we face a non-trivial problem unless we make some
simplifying assumptions, which, however, will still introduce DSR
corrections pertaining to the Planck scale cutoff.

As the spherical symmetry remains intact in the DSR theory
\cite{sg} we define the respective components of energy-momentum
tensor $\tilde{t}^{\mu \nu}$ in the NC framework analogous to
(\ref{lD}, \ref{lN}) as,
\begin{equation}
\tilde{t}^{ii} = p = \frac{1}{3}
\sum_{i}\frac{\bf{p}_{i}^{2}}{p^0_{i}} \delta^{3} \left( x -
x_i\right);
$$$$
\tilde{t}^{00}=\rho = \sum_{i} p^0_i \delta^{3}\left( x -
x_i\right) ~~;~~ n = \sum_{i}\delta^{3}\left( x - x_i\right),
\label{lr}
\end{equation}
where $\bf{p}_i$ and $p^0_i$ are respectively the momentum
three-vector and the energy of the $i$-th fluid particle in the
DSR spacetime. Using (\ref{ld}) and using the scaling properties
of Dirac-$\delta$ function we obtain the following results,
\begin{equation}
F^{-1}(p) = \frac{1}{3} \sum_{i}\frac{\bf{P}_{i}^{2}}{P^0_{i} (1 +
P^0_i/\kappa)^4} \delta^{3} \left( X - X_i\right), \label{le}
\end{equation}
\begin{equation}
F^{-1}(\rho) = \sum_{i} \frac{P^0_i}{(1 + P^0_i/\kappa)^4}
\delta^{3}\left( X - X_i\right), \label{lf}
\end{equation}
\begin{equation}
F^{-1}(n) = \sum_{i}
\frac{N}{\left(1+\frac{P^0_i}{\kappa}\right)^3}\delta^{3}\left( X
- X_i\right).
\end{equation}
In a combined form, we can write down the following non-linear
mapping (and its inverse) as,
\begin{equation}
F^{-1}(\tilde{t}^{\mu\nu})=\sum_{i}\frac{P^\mu_{i}
P^\nu_{i}}{P^0_{i} (1 + P^0_i/\kappa)^4} \delta^{3} \left( X -
X_i\right)~~~,~~~ F(\tilde{T}^{\mu\nu})=\sum_{i}\frac{p^\mu_{i}
p^\nu_{i}}{p^0_{i} (1 + p^0_i/\kappa)^4} \delta^{3} \left( x -
x_i\right). \label{ft}
\end{equation}
The way we have defined the DSR EMT it is clear that comoving form
of EMT also receives DSR corrections. But problem crops up when,
in analogy to SR EMT \cite{wein}, we attempt to boost the $\tilde
t^{\mu\nu}$ to a laboratory frame with an arbitrary velocity
$v^i$. Recall that for a single particle DSR boosts involve its
energy and momentum. Since $p$ and $\rho$ (for $\tilde
t^{\mu\nu}$) denote composite variables it is not clear which
energy or momentum will come into play. To proceed further in the
DSR boost we put in a single energy $\bar p^0$ and momentum $\bar
p^i$ that signifies the average energy and momentum (modulus) of
the whole fluid. In fact this simplification is not very
artificial since we are obviously considering equilibrium systems
(however see \cite{newref}) . This allows us to use the mappings:
\begin{equation}
F^{-1}(p) = \frac{P}{(1+\bar P^0/\kappa)^4}~~,~~F^{-1}(\rho) =
\frac{D}{(1+\bar P^0/\kappa)^4}~~,~~F^{-1}(n) = \frac{N}{(1+\bar
P^0/\kappa)^4}.
\end{equation}
In a covariant form the mapping and its inverse between $\tilde{t}^{\mu
\nu}$ and $\tilde{T}^{\mu \nu}$ are,
\begin{equation}
F^{-1}(\tilde{t}^{\mu \nu}) = \frac{\tilde{T}^{\mu
\nu}}{(1+\bar P^0/\kappa)^4}~~~~,~~~~F(\tilde{T}^{\mu \nu}) =
\frac{\tilde{t}^{\mu \nu}}{(1-\bar p^0/\kappa)^4}. \label{ftf}
\end{equation}
Finally we can apply the definition of $L_{DSR}$ using (\ref{ftf})
with (\ref{lct}) to obtain the following expressions for
energy-momentum tensor with respect to an arbitrary inertial frame
in a DSR spacetime,
\begin{equation}
t^{00}=L_{DSR}(\tilde{t}^{00}) = F \circ L_{SR} \circ F^{-1}
(\tilde{t}^{00}) = F \circ L_{SR}
\left(\frac{\tilde{T}^{00}}{(1+\bar P^0/ \kappa)^4}\right) $$$$ =
F \left(\frac{\gamma^2(D+P
v^2)}{\left(1+\frac{\gamma}{\kappa}\left(\bar P^0-v\bar
P^1\right)\right)^4}\right) = \frac{\gamma^2(\rho+p v^2)}{\bar
\alpha^4};
$$$$
t^{i0}=L_{DSR}(\tilde{t}^{i0}) = \frac{\gamma^2(\rho+p )v^i}{\bar
\alpha^4}~~;~~ t^{ij}=L_{DSR}(\tilde{t}^{ij}) =
\frac{\gamma^2(\rho+p )v^i v^j}{\bar \alpha^4}+p \delta^{ij}.
\label{dsrmt}
\end{equation}
It is very interesting to note that the above expressions can also
be combined into a single form which is structurally very close to
the fluid EMT in SR,
\begin{equation}
t^{\mu\nu}=\frac{\left(1-\frac{\bar p_0}{\kappa}\right)^2}{\bar
\alpha^{4}} \left((p+\rho)u^{\mu}u^{\nu}+
p\frac{\eta^{\mu\nu}}{\left(1-\frac{\bar
p_0}{\kappa}\right)^2}\right)= \frac{\left(1-\frac{\bar
p_0}{\kappa}\right)^2}{\bar \alpha^{4}}
\left((p+\rho)u^{\mu}u^{\nu}+ p\tilde{\eta}^{\mu\nu}\right).
\label{lnctf}
\end{equation}
where we have defined the four-velocity $u^\mu$ in the DSR
spacetime as:
\begin{equation}
u^0 = dx^0/d\tau = \frac{\gamma}{\left(1-\bar
p_0/\kappa\right)}~~~,~~~u^i = dx^i/d\tau = \frac{\gamma
v^i}{\left(1-\bar p_0/\kappa\right)}.
\end{equation}
Note that the DSR four-velocity $u^\mu$ is actually the mapped
form of the SR four-velocity $U^\mu$ since the parameter $\tau $
does not undergo any transformation. The other point to notice is
that $\tilde\eta ^{\mu\nu }$ of (\ref{ncm}), (DSR analogue of the
flat metric $\tilde\eta ^{\mu\nu }$), appears in $t^{\mu\nu }$
making the final form of the DSR EMT transparent. Indeed
$t^{\mu\nu }$ in (\ref{lnctf}) reduces smoothly to $T^{\mu\nu }$
of SR (\ref{lctf}) in the large $\kappa$ limit. Incidentally,
again in analogy to the SR construction of many-body system for
fluid ((13), (14)), this form of $t^{\mu\nu }$ is consistent with
the microscopic picture of DSR EMT for fluid that we have
developed ((19), (20), (21), (22), (23)).

\section {\bf {Equation of state}}
So far we have only provided the abstract form of DSR EMT,
relevant for a fluid, from purely kinematical considerations. It
is now time for application. Keeping an eye in our cosmological
motivation, in the present paper we will take up the issue of
equation of state for an ideal DSR fluid.

\subsection {\bf {Equation of state in SR theory:}}~~~In the standard SR
version, one way of deriving \cite{wein} the equation of state is
to return to the microscopic picture (\ref{lt1}) and substitute
the SR energy dispersion relation $P^0 = ({\bf{P}}^2 + m^2)^{1/2}$
into (\ref{lD}) to get the following expression for the equation
of state,
\begin{equation}
P=\frac{1}{3}D - \frac{1}{3} \sum_{i} \frac{m^2}{P^0_i}
\delta^3(X-X_i) \label{PD}.
\end{equation}
For a cool non-relativistic gas we have ${\bf{P}}\ll m$; so the
expression for the energy becomes:~~$E \simeq m +
\frac{{\bf{P}}^2}{2 m}$. Using (\ref{lD}) and (\ref{lN}) one gets
the equation of state
\begin{equation}
D-mN=\frac{3}{2}P \label{nrPD}.
\end{equation}
For a hot ultra-relativistic gas since~ $E \simeq {\bf{P}} \gg m$
using (\ref{lD}) the equation of state becomes
\begin{equation}
D=3P \label{urPD}.
\end{equation}

\subsection {\bf {Equation of state in DSR theory:}}~~~
Let us now we proceed to derive the ideal fluid equation of state
in the DSR scheme. We start with the Magueijo-Smolin modified
dispersion relation (\ref{ms}),
\begin{equation}
(\bar{p}^0)^2 - {\bf{\bar{p}}}^2 =
m^2\left[1-\frac{\bar{p}^0}{\kappa}\right]^2.
\label{ms0}
\end{equation}
We solve this
equation for $\bar{p}^0$ to $O(\kappa )$,
\begin{equation}
\bar{p}^0= \left({\bf{\bar{p}}}^2+m^2\right)^\frac{1}{2} -
\frac{m^2}{\kappa}.
\end{equation}
We substitute this expression in (\ref{lr}) and finally obtain,
\begin{equation}
p=\frac{1}{3}\rho - \frac{1}{3} \sum_{i} \frac{m^2}{\bar{p}^0}
\delta^3(x-x_i) + \frac{2m^2n}{\kappa} \label{pr}.
\end{equation}
In the non-relativistic regime $\bar{p}^0 \simeq m +
\frac{{\bf{\bar{p}}}^2}{2m}-\frac{m^2}{\kappa}$, using (\ref{lr}) we
have
\begin{equation}
\rho-mn=\frac{3}{2} p - \frac{m^2 n}{\kappa} \label{nrpr}.
\end{equation}
However something interesting occurs in the extreme relativistic
scenario due to the Planck energy upper bound $\bar{p}^0\sim
\kappa $. Referring once again to the Magueijo-Smolin dispersion
relation (\ref{ms0}), we find that for $\bar{p}^0=\kappa $ the SR
photon dispersion relation is recovered, $\bar{p}^0=\mid \bf
\bar{p} \mid =\kappa $ the rest mass $m$ does not appear in the
consideration. (In fact one can check that the energy ceiling
$\kappa$ can only be reached by a massless particle.) But this
condition reduces the equation of state to,
\begin{equation}
\rho=3p=n\kappa. \label{urpr}
\end{equation}
These
equations of state might prove to be important signatures for
quantum gravity effects if DSR happens to be the proper framework
to address quantum gravity issues.

\section {\bf {Conclusion and future prospects}}
Doubly Special Relativity (DSR) is a generalization of Special Relativity
(SR) that can be relevant in the context
of quantum gravity since it possesses an observer invariant energy upper
bound, naturally assumed to be the Planck energy. Also DSR is compatible with the $\kappa$-Minkowski form of noncommutative spacetime. DSR reduces to
SR for low energy regime as indeed it should. In this paper, for the first time,
we have tried to incorporate DSR effects in an ideal fluid since
eventually we aim to consider a DSR based cosmology.

We  generalize  the conventional framework of deriving the covariant energy-momentum tensor by  boosting its spherically symmetric form, where we exploit the DSR-Lorentz transformations (instead of the Special Theory transformations).
 We stress that effects of a noncommutative (in particular $\kappa$-Minkowski) spacetime enters through the DSR-Lorentz transformations.
In the process we had to resort to some simplifying assumptions
in describing the fluid as a many-body system
(in the so called Lagrangian description of fluid).
One might treat this problem as a more virulent form of the one
we find even in SR if we try to treat a multi-particle system in a relativistic way.
We have exploited the concept that DSR is a non-linear realization of SR so
that one can use a canonical phase space as a tool for obtaining DSR relations.
We have demonstrated that, even in this simplified situation,
there are effects of DSR that introduces the Planck scale in the
equations of motion for an ideal fluid. Below we list some of
the open problems that we plan to pursue in near future: \\
(I) While boosting the comoving form of energy momentum tensor in DSR, we had to
utilize the average values of energy and momentum modulus for the
whole system while (DSR) boosting. We require an improved way of
applying the DSR boost keeping the dependence of DSR boost on
individual particles of the fluid intact.\\
(II) Two of us are looking at the thermodynamics of ideal fluid for
DSR explicitly from the partition function \cite{sddr}.
In this formulation DSR effects will appear from two sources,
from the deformed mass-energy dispersion relation of particles and
from the high energy cut off in the form of Planck energy.\\
(III) Generalization of Cosmology in DSR framework is the next
programme that we wish to take up. \vskip .5cm
{\bf{Acknowledgements:}} One of us (DR) wishes to thank Dr. A.
Mukherjee for discussions. He is also grateful to Physics and
Applied Mathematics Unit, Indian Statistical Institute, where most
of this work was done, and to C.S.I.R., India, for financial
support.

\end{document}